\documentclass[conference]{IEEEtran}
\usepackage{multirow}
\usepackage{diagbox}
\usepackage{amssymb}
\usepackage{amsmath}
\usepackage{graphicx}
\usepackage{cite}
\usepackage{citesort}
\usepackage{subfigure}
\usepackage{graphicx,epstopdf}
\usepackage{epsfig}	
\usepackage{cite,graphicx,amsmath,amssymb}
\usepackage{comment}

\usepackage{amssymb}
\usepackage{amsmath}
\usepackage{cite}
\usepackage{url}
\usepackage{xcolor}
\usepackage{cite,graphicx,amsmath,amssymb}
\usepackage{subfigure}
\usepackage{citesort}
\usepackage{fancyhdr}
\usepackage{mdwmath}
\usepackage{mdwtab}
\usepackage{caption}
\usepackage{amsthm}
\usepackage{lipsum}

\newtheorem{theorem}{Theorem}

\newtheorem{lemma}{Lemma}

\newtheorem{corollary}{Corollary}

\newtheorem{remark}{Remark}  

\makeatletter
\def\ScaleIfNeeded{%
\ifdim\Gin@nat@width>\linewidth \linewidth \else \Gin@nat@width
\fi } \makeatother

\begin{document}

\title{Outage Performance of A Unified Non-Orthogonal Multiple Access Framework}

\author{
\IEEEauthorblockN{  Xinwei~Yue\IEEEauthorrefmark{1}, Zhijin Qin\IEEEauthorrefmark{2}, Yuanwei Liu\IEEEauthorrefmark{3}, Xiaoming Dai\IEEEauthorrefmark{4}, and Yue Chen\IEEEauthorrefmark{3}   }
\IEEEauthorblockA{\IEEEauthorrefmark{1} Beihang University, Beijing, China\\
\IEEEauthorrefmark{2} Lancaster University, Lancaster, UK\\
\IEEEauthorrefmark{3} Queen Mary University of London, London, UK\\
\IEEEauthorrefmark{4} University of Science and Technology Beijing, Beijing, China
 } }



\maketitle

\begin{abstract}
In this paper, a unified framework of non-orthogonal multiple access (NOMA) networks is proposed, which can be applied to code-domain NOMA (CD-NOMA) and power-domain NOMA (PD-NOMA).
Since the detection of NOMA users mainly depend on efficient successive interference cancellation (SIC) schemes, both imperfect SIC (ipSIC) and perfect SIC (pSIC) are taken into considered. To characterize the performance of this unified framework, the exact and asymptotic expressions of outage probabilities as well as delay-limited throughput for CD/PD-NOMA with ipSIC/pSIC are derived. Based on the asymptotic analysis, the diversity orders of CD/PD-NOMA are provided. It is confirmed that due to the impact of residual interference (RI), the outage probability of the $n$-th user with ipSIC for CD/PD-NOMA converges to an error floor in the high signal-to-noise ratio (SNR) region. Numerical simulations demonstrate that the outage behavior of CD-NOMA is superior to that of PD-NOMA.
\end{abstract}


\section{Introduction}
To enhance spectrum efficiency and massive connectivity, non-orthogonal multiple access (NOMA) \cite{QinNOMA,Yuanwei2017Proceeding} has been identified as one of the key technologies for the fifth generation (5G) networks.
The pivotal feature of NOMA is its capability of sharing the same physical resource element (RE), where multiple users' signals are linearly superposed over different power levels by using the superposition coding scheme. To get the desired signal, multi-user detection algorithm \cite{Cover1991Elements}, i.e., successive interference cancellation (SIC) or message passing algorithm (MPA) is carried out at the receiver.

More particularly, based on spreading signature of multiple access (MA), NOMA schemes can be divided into two categories: power-domain NOMA (PD-NOMA) and code-power NOMA (CD-NOMA). In \cite{Ding2014performance}, two evaluation metrics of PD-NOMA networks including outage probability and ergodic rate have been proposed, where the outage behaviors of users and ergodic rate have been discussed by applying stochastic geometry. From a practical perspective, the authors in \cite{Yang7361990} studied the performance of PD-NOMA for the two-user case with imperfect channel state information (CSI), where the closed-form expressions of outage probability were derived. When NOMA users have similar channel conditions, the authors of \cite{Ding7906532Multicast} proposed a PD-NOMA based multicast-unicast scheme and verified that the spectral efficiency of PD-NOMA based multicast-unicast scheme is higher than that of orthogonal multiple access (OMA) based one.

As a further advance, CD-NOMA is viewed as a special extension of PD-NOMA, in which the data streams of multiple users are directly mapped into multiple REs (or subcarriers $K$) through the sparse matrix/codebook or low density spread sequence. Actually, CD-NOMA mainly include sparse code multiple access (SCMA), pattern division multiple access (PDMA), multi-user sharing access (MUSA), etc.
In \cite{Nikopour2013Sparse}, the authors proposed a sub-optimal design approach to design the sparse codebook of SCMA. On the condition of the fixed sparse pattern matrix, the authors of \cite{ChenPattern7526461} evaluated the link level performance of PDMA and confirmed that PDMA can achieve the higher spectrum efficiency than OMA. In \cite{Yuan2016Multi}, MUSA is capable of adopting a grant-free scheme to support Internet of Things (IoT) scenario. However, up to now, there is no work investigating the performance of the unified NOMA framework.

Driven by this, we investigate the outage performance of the unified NOMA framework by invoking stochastic geometry. Since the detection of NOMA users mainly depend on efficient SIC schemes, both imperfect SIC (ipSIC) and perfect SIC (pSIC) are taken into considered. We derive the exact expressions of outage probability for a pair of NOMA users (i.e., the $n$-th user and $m$-th user) in the unified framework. To obtain deep insights, we further derive the asymptotic outage probability of two users and attain the corresponding diversity orders. Due to the impact of residual interference (RI), the outage behavior of the $n$-th user with ipSIC for CD/PD-NOMA (CD-NOMA and PD-NOMA) converges to an error floor. Furthermore, we confirm that the outage behavior of CD-NOMA is superior to that of PD-NOMA.
Additionally, we analyze system throughput for CD/PD-NOMA in the delay-limited transmission.

\section{Network Model}\label{Network Model}
\subsection{Network Descriptions}
Consider a unified NOMA downlink transmission scenario, where a base station (BS) transmits the information to $M$ randomly users. The BS directly maps the data streams of multiple users into $K$ subcarriers or REs by utilizing one sparse spreading matrix ${{\bf{G}}_{K \times M}}$, in which there are a few number of non-zero entries within it and satisfies the relationship $M > K$. For simplicity, we assume that the BS and NOMA users are equipped with a single antenna, respectively. Assuming that the BS is located at the center of circular cluster denoted as $D$, with radius $R_D$ and the $M$ NOMA users are uniformly distributed within  circular cluster \cite{Yuanwei2017TWC}.
To facilitate analysis, we assumed that $M$ users are divided into $M/2$ orthogonal pairs, in which the distant user and the nearby user can be distinguished based on their disparate channel conditions. Furthermore, each pair of users is randomly selected to carry out the NOMA protocol \cite{Ding2014performance,Yuanwei2017JSAC}. A bound pass model is employed to model the channel coefficients in networks from the BS to users. Meanwhile, these wireless links are disturbed by additive white Gaussian noise (AWGN) with mean power ${{N_0}}$. Without loss of generality, the effective channel gains between the BS and users are sorted as $\left\| {{{\bf{h}}_M}} \right\|_2^2 >  \cdots  > \left\| {{{\bf{h}}_n}} \right\|_2^2 >  \cdots  > \left\| {{{\bf{h}}_m}} \right\|_2^2 >  \cdots  > \left\| {{{\bf{h}}_1}} \right\|_2^2$ \cite{David2003Order,Men7219393} with the assistance of order statistics. In this paper, we focus on the $m$-th user paired with the $n$-th user for NOMA transmission.
\subsection{Signal Model}
Regarding the unified NOMA downlink transmission scenario, the BS transmits the superposed signals to multiple users, where the data stream of each user spreads over one column of sparse matrix. Hence the observation at the $\varphi $-th user over $K$ subcarriers ${{\bf{y}}_\varphi } = {[{y_{\varphi 1}}{y_{\varphi 2}} \cdots {y_{\varphi K}}]^T}$ is given by
\begin{align}\label{The received signal expression at the n and m-th user}
{{\bf{y}}_\varphi }{\rm{ = }}diag\left( {{{\bf{h}}_\varphi }} \right) ( {{{\bf{g}}_n}\sqrt {{P_s}{a_n}} {x_n} + {{\bf{g}}_m}\sqrt {{P_s}{a_m}} {x_m}} ) + {{\bf{n}}_\varphi },
\end{align}
where $\varphi  \in \left( {n,m} \right)$, $x_{n}$ and $x_{m}$ are the normalized unity power signals for the $n$-th and $m$-th users, respectively, i.e, $\mathbb{E}\{x_{n}^2\}= \mathbb{E}\{x_{m}^2\}=1$.
We assume the fixed power allocation coefficients satisfy the condition that ${a_m} > {a_n}$ with $a_m + a_n = 1$, which is for fairness considerations. $P_{s}$ denotes the normalized transmission power at BS. The sparse indicator vector of the $\varphi $-th user is denoted by ${{\bf{g}}_\varphi } = {[{g_{\varphi 1}}{g_{\varphi 2}} \cdots {g_{\varphi K}}]^T}$, which is one column of ${{\bf{G}}_{K \times M}}$. More specifically, ${g_{\varphi k}}$ is the subcarrier index, where ${g_{\varphi k}}=1$ indicates the signals are mapped into the corresponding RE, otherwise, ${g_{\varphi k}}=0$.
Let ${{\bf{h}}_\varphi } = {[{{\tilde h}_{\varphi 1}}{{\tilde h}_{\varphi 2}} \cdots {{\tilde h}_{\varphi K}}]^T}$ denotes the channel vector between the BS and $\varphi $-th user occupying $K$ subcarriers with ${{\tilde h}_{\varphi k}} = \frac{{\sqrt \eta  {h_{\varphi k}}}}{{\sqrt {1 + {d^\alpha }} }}$, where ${h_{\varphi k}} \sim {\cal C}{\cal N}\left( {0,1} \right)$ is the Rayleigh fading channel gain between the BS and $\varphi $-th user occupying the $k$-th subcarrier,
$\eta$ is a frequency dependent factor, $\alpha $ is the path loss exponent and $d$ is the distance from BS to $\varphi $-th user.
${{\bf{n}}_\varphi } \sim {\cal C}{\cal N}\left( {0,{N_0}{{\bf{I}}_{K}}} \right)$ denotes the AWGN.

To maximize the output SNRs and diversity orders, we employ the maximal ratio combiner (MRC) at the $\varphi $-th user over $K$ subcarriers. Let ${{\mathbf{u}}_\varphi } = \frac{{{{\left( {diag\left( {{{\mathbf{h}}_\varphi }} \right){{\mathbf{g}}_\varphi }} \right)}^*}}}{{\left\| {diag\left( {{{\mathbf{h}}_\varphi }}\right){{\mathbf{g}}_\varphi }} \right\|}}$, and then the received signal at the $\varphi $-th user can be written as
\begin{align}\label{The further received signal expression at the n and m-th user}
{{\tilde y}_\varphi }{\text{ = }}{{\mathbf{u}}_\varphi }diag\left( {{{\mathbf{h}}_\varphi }} \right) ( {{{\mathbf{g}}_n}\sqrt {{P_s}{a_n}} {x_n} + {{\mathbf{g}}_m}\sqrt {{P_s}{a_m}} {x_m}} ) + {{\mathbf{u}}_\varphi }{{\mathbf{n}}_\varphi }.
\end{align}
Based on aforementioned assumptions, the signal-plus-interference-to-noise ratio (SINR) at the $n$-th user to detect the $m$-th user's signal $x_{m}$ is given by
\begin{align}\label{The SINR expression at the n-th user to detect the m-th user}
{\gamma _{n \to m}} = \frac{{\rho \left\| {diag\left( {{{\bf{h}}_n}} \right){{\bf{g}}_m}} \right\|_2^2{a_m}}}{{\rho \left\| {diag\left( {{{\bf{h}}_n}} \right){{\bf{g}}_n}} \right\|_2^2{a_n} + 1}},
\end{align}
where $\rho  = \frac{{{P_s}}}{{{N_0}}}$ denotes the transmit SNR. For the sake of simplicity, assuming that ${{{\bf{g}}_m}}$ and ${{{\bf{g}}_n}}$ have the same column weights for ${{\bf{G}}_{K \times M}}$.

By applying SIC~\cite{Zhiguo2015Mag}, the SINR of the $n$-th user, who needs to decode the information of itself is given by
\begin{align}\label{the SINR expression at n-th user to detect itself with SIC}
{\gamma _n} = \frac{{\rho \left\| {diag\left( {{{\bf{h}}_n}} \right){{\bf{g}}_n}} \right\|_2^2{a_n}}}{{\varpi \rho \left\| {{{\bf{h}}_I}} \right\|_2^2 + 1}},
\end{align}
where $\varpi {\rm{ = 0}}$ and $\varpi {\rm{ = 1}}$ denote the pSIC and ipSIC operations, respectively. Note that
${{\bf{h}}_I} = {[{h_{I1}}{h_{I2}} \cdots {h_{IK}}]^T}$ denotes the RI channel vector at $K$ subcarriers with $ {h_{Ik}} \sim {\cal C}{\cal N}\left( {0,{\Omega _I}} \right)$.

The SINR of the $m$-th NOMA user to decode the information of itself can be expressed as
\begin{align}\label{SINR m}
{\gamma _m} = \frac{{\rho \left\| {diag\left( {{{\bf{h}}_m}} \right){{\bf{g}}_m}} \right\|_2^2{a_m}}}{{\rho \left\| {diag\left( {{{\bf{h}}_m}} \right){{\bf{g}}_n}} \right\|_2^2{a_n} + 1}}.
\end{align}
\section{Performance evaluation}\label{Connection Outage}
In this section, the outage probability for a pair of NOMA users is selected as a metric to evaluate the performance of the unified downlink NOMA networks.
\subsection{The outage probability of the $m$-th user}
The outage event of the $m$-th user is that the $m$-th user cannot detect its own information. Hence the outage probability of the $m$-th user for CD-NOMA can be expressed as
\begin{align}\label{CD-NOMA:the expression of COP for far user}
{P_{m,CD}} = {\rm{Pr}}\left( {{\gamma _m} < {\varepsilon _m}} \right),
\end{align}
where ${\varepsilon _m} = {2^{{R_m}}} - 1$ and $R_m$ is the target rate of the $m$-th user.
The following theorem provides the outage probability of the $m$-th user.
\begin{theorem}\label{Theorem:CD-NOMA:the COP of far user}
The outage probability of the $m$-th user for CD-NOMA is given by
\begin{align}\label{CD-NOMA:the COP of far user}
{P_{m,CD}} \approx & {\phi _m}\sum\limits_{p = 0}^{M - m} {{
 M - m \choose
 p
 }\frac{{{{\left( { - 1} \right)}^p}}}{{m + p}}}  \nonumber \\
 & \times {\left[ {\sum\limits_{u = 1}^U {{b_u}} \left( {1 - {e^{ - \frac{{\tau {c_u}}}{\eta }}}\sum\limits_{i = 0}^{K - 1} {\frac{1}{{i!}}{{\left( {\frac{{\tau {c_u}}}{\eta }} \right)}^i}} } \right)} \right]^{m + p}}  ,
\end{align}
where $\tau  = \frac{{{\varepsilon _m}}}{{\rho \left( {{a_m} - {\varepsilon _m}{a_n}} \right)}}$ with ${a_m} > {\varepsilon _m}{a_n}$, ${\phi _m} = \frac{{M!}}{{\left( {M - m} \right)!\left( {m - 1} \right)!}}$, ${
 M - m \choose
 p
 } = \frac{{\left( {M - m} \right)!}}{{p!\left( {M - m - p} \right)!}}$, ${b_u} = \frac{\pi }{{2U}}\sqrt {1 - \theta _u^2} \left( {{\theta _u} + 1} \right)$, ${c_u}{\rm{ = }}1 + {\left( {\frac{{{R_D}}}{2}\left( {{\theta _u} + 1} \right)} \right)^\alpha }$, ${\theta _u} = \cos \left( {\frac{{2u - 1}}{{2U}}\pi } \right)$ and $U$ is a parameter to ensure a complexity-accuracy tradeoff.
 \begin{proof}
See Appendix~A.
\end{proof}
\end{theorem}

\begin{corollary}\label{Corollary:PD-NOMA:the COP of far user}
For the special case with $K=1$, the outage probability of the $m$-th user for PD-NOMA is given by
\begin{align}\label{PD-NOMA:the COP of far user}
{P_{m,PD}} \approx {\phi _m}\sum\limits_{p = 0}^{M - m} {{
 M - m \choose
 p
}\frac{{{{\left( { - 1} \right)}^p}}}{{m + p}}} {\left[ {\sum\limits_{u = 1}^U {{b_u}} \left( {1 - {e^{ - \frac{{{{{ \tau }}} {c_u}}}{\eta }}}} \right)} \right]^{m + p}}.
\end{align}
\end{corollary}
\subsection{The outage probability of the $n$-th user}
As stated in \cite{Ding2014performance,Liu7445146SWIPT}, the outage for the $n$-th user can happen in the following two cases : 1) The $n$-th user cannot decode the message of the $m$-th user; and 2) The $n$-th user can decode the message of the $m$-th user, then carries out SIC operations, but cannot decode the information of itself. Hence the outage probability of the $n$-th user can be expressed as
\begin{align}\label{CD-NOMA:the expression of COP for near user}
P_{n,CD} =& \Pr \left\{ {{\gamma _{n \to m}} \le {\varepsilon _m}} \right\} \nonumber \\
&+ \Pr \left\{ {{\gamma _{n \to m}} > {\varepsilon _m},{\gamma _n} \le {\varepsilon _n}} \right\},
\end{align}
where ${\varepsilon _n} = {2^{{R_n}}} - 1$ with ${R_n}$ being the target rate at the $n$-th user to detect the $m$-th user.
The following theorem provides the outage probability of the $n$-th user with ipSIC for CD-NOMA.
\begin{theorem}\label{Theorem:CD-NOMA:the COP of near user for Case1 with ipSIC}
The outage probability of the $n$-th user with ipSIC for CD-NOMA is given by \eqref{CD-NOMA:the COP of near user for Case1 with ipSIC}, where $\beta {\text{ = }}\frac{{{\varepsilon _n}}}{{\rho {a_n}}}$, $\vartheta {\text{ = }}\frac{{\varpi {\varepsilon _n}}}{{{a_n}}}$ and $\varpi {\text{ = }}1$.
\begin{figure*}[!t]
\normalsize
\begin{align}\label{CD-NOMA:the COP of near user for Case1 with ipSIC}
P_{n,CD}^{ipSIC} \approx \frac{{{\phi _n}}}{{\left( {K - 1} \right)\Omega _I^K}}\sum\limits_{p = 0}^{M - n} {{
   {M - n}  \choose
   p
} \frac{{{{\left( { - 1} \right)}^p}}}{{n + p}}} \int_0^\infty  {{y^{K - 1}}{e^{ - \frac{y}{{{\Omega _I}}}}}} {\left[ {\sum\limits_{u = 1}^U {{b_u}} \left( {1 - {e^{ - \frac{{{c_u}\left( {\vartheta y + \beta } \right)}}{\eta }}}\sum\limits_{i = 0}^{K - 1} {\frac{1}{{i!}}{{\left( {\frac{{\left( {\vartheta y + \beta } \right){c_u}}}{\eta }} \right)}^i}} } \right)} \right]^{n + p}}dy.
\end{align}
\hrulefill \vspace*{0pt}
\end{figure*}
\begin{proof}
See Appendix~B.
\end{proof}
\end{theorem}

Substituting $\varpi {\text{ = }}0$ into \eqref{CD-NOMA:the COP of near user for Case1 with ipSIC}, the outage probability of the $n$-th user with pSIC for CD-NOMA is given by
\begin{align}\label{CD-NOMA:the COP of near user for Case1 with pSIC}
P_{n,CD}^{pSIC} \approx & {\phi _n}\sum\limits_{p = 0}^{M - n} {{
   {M - n}  \choose
   p
 }\frac{{{{\left( { - 1} \right)}^p}}}
{{n + p}}\left[ {\sum\limits_{u = 1}^U {{b_u}} } \right.}  \hfill  \nonumber \\
  &{\left. { \times \left( {1 - {e^{ - \frac{{\beta {c_u}}}
{\eta }}}\sum\limits_{i = 0}^{K - 1} {\frac{1}
{{i!}}{{\left( {\frac{{\beta {c_u}}}
{\eta }} \right)}^i}} } \right)} \right]^{n + p}} \hfill .
\end{align}

\begin{corollary}\label{Corollary:PD-NOMA:the COP of near user for case1}
For the special case with $K=1$, the outage probability of the $n$-th user with ipSIC for PD-NOMA is given by
\begin{align}\label{PD-NOMA :the COP of near user with ipSIC for Case1}
P_{n,PD}^{ipSIC} \approx& \frac{{{\phi _n}}}{{{\Omega _I}}}\sum\limits_{p = 0}^{M - n} {{
   {M - n}  \choose
   p
}\frac{{{{\left( { - 1} \right)}^p}}}{{n + p}}} \nonumber  \\
&  \times \int_0^\infty  {{e^{ - \frac{y}{{{\Omega _I}}}}}} {\left[ {\sum\limits_{u = 1}^U {{b_u}} \left( {1 - {e^{ - \frac{{{c_u}\left( {\vartheta y + \beta } \right)}}{\eta }}}} \right)} \right]^{n + p}}dy  .
\end{align}
\end{corollary}

Substituting $\varpi {\text{ = }}0$ into \eqref{PD-NOMA :the COP of near user with ipSIC for Case1}, the outage probability of the $n$-th user with pSIC for PD-NOMA is given by
\begin{align}\label{PD-NOMA:the COP of near user with pSIC for Case1}
P_{n,PD}^{pSIC} \approx & {\phi _n}\sum\limits_{p = 0}^{M - n} {{
   {M - n}  \choose
   p
 }\frac{{{{\left( { - 1} \right)}^p}}}
{{n + p}}\left[ {\sum\limits_{u = 1}^U {{b_u}} } \right.}  \hfill \nonumber \\
  & {\left. { \times \left( {1 - {e^{ - \frac{{ \beta {c_u}}}
{\eta }}}} \right)} \right]^{n + p}} \hfill .
\end{align}

\subsection{Diversity Order Analysis}
To obtain deep insights, diversity order analysis is present, which highlights the slope of the curves for outage probabilities varying with the SNRs. The definition of diversity order is given by
\begin{align}\label{The diversity order of COP}
d =  - \mathop {\lim }\limits_{\rho  \to \infty } \frac{{\log \left( {P^\infty }\left( \rho  \right) \right)}}{{\log \rho }},
\end{align}
where ${P^\infty }\left( \rho  \right)$ denotes the asymptotic outage probability at high SNR region.

\begin{corollary}\label{Corollary:the expression of asymptotic COP for the m-th user}
The asymptotic outage probability of the $m$-th user for CD-NOMA is given by
\begin{align}\label{The expression of asymptotic COP for the m-th user with CD-NOMA}
P_{m,CD}^\infty  \approx \frac{{M!}}
{{\left( {M - m} \right)!m!}}{\left[ {\sum\limits_{u = 1}^U  \frac{{{b_u}}}
{{K!}}{{\left( {\frac{{\tau {c_u}}}
{\eta }} \right)}^K}} \right]^m}.
\end{align}
\begin{proof}
By definition, ${\Theta _1}{\text{ = }}1 - {e^{ - \frac{{\tau {c_u}}}{\eta }}}\underbrace {\sum\limits_{i = 0}^{K - 1} {\frac{1}
{{i!}}{{\left( {\frac{{\tau {c_u}}}{\eta }} \right)}^i}} }_{{\Theta _2}}$. Applying power series expansion, the summation term ${\Theta _2}$ can be rewritten as ${\Theta _2} = {e^{\frac{{\tau {c_u}}} {\eta }}} - \sum\limits_{i = K}^\infty  {\frac{1}
{{i!}}{{\left( {\frac{{\tau {c_u}}}{\eta }} \right)}^i}} $.  Substituting ${\Theta _2}$ into ${\Theta _1}$, when $x  \to 0 $, ${\Theta _1}$ with the approximation of ${e^{ - x}} \approx 1 - x$ is formulated as ${\Theta _1} \approx \frac{1}{{K!}}{\left( {\frac{{\tau {c_u}}}{\eta }} \right)^K}$. Furthermore, substituting ${\Theta _1}$ into \eqref{CD-NOMA:the COP of far user} and taking the first term $(p=0)$, we obtain \eqref{The expression of asymptotic COP for the m-th user with CD-NOMA}. The proof is completed.
\end{proof}
\end{corollary}

For the special case with $K=1$, the asymptotic outage probability of the $m$-th user for PD-NOMA is given by
\begin{align}\label{The expression of asymptotic COP for the m-th user with PD-NOMA}
P_{m,PD}^\infty  \approx \frac{{M!}}
{{\left( {M - m} \right)!m!}}{\left[ {\sum\limits_{u = 1}^U {{b_u}} \left( {\frac{{\tau {c_u}}}
{\eta }} \right)} \right]^m}.
\end{align}
\begin{remark}\label{remark1: the m-th user}
Upon substituting \eqref{The expression of asymptotic COP for the m-th user with CD-NOMA} and \eqref{The expression of asymptotic COP for the m-th user with PD-NOMA} into \eqref{The diversity order of COP}, the diversity orders of the $m$-th user for CD-NOMA and PD-NOMA are $mK$ and $m$, respectively.
\end{remark}
\begin{corollary}\label{Corollary:the expression of asymptotic COP for the n-th user CD-NOMA case1}
The asymptotic outage probability of the $n$-th user with ipSIC for CD-NOMA is given by
\begin{align}\label{The asymptotic expression of the n-th user with ipSIC for CD-NOMA case1}
P_{n,CD}^{ipSIC,\infty } \approx \frac{{{\phi _n}}}{{\left( {K - 1} \right)\Omega _I^K}}\sum\limits_{p = 0}^{M - n} {{
   {M - n}  \choose
   p
}\frac{{{{\left( { - 1} \right)}^p}}}{{n + p}}} \int_0^\infty  {{y^{K - 1}}}  \nonumber \\
 \times {e^{ - \frac{y}{{{\Omega _I}}}}}{\left[ {\sum\limits_{u = 1}^U {{b_u}} \left( {1 - {e^{ - \frac{{y\vartheta {c_u}}}{\eta }}}\sum\limits_{i = 0}^{K - 1} {\frac{1}{{i!}}{{\left( {\frac{{y\vartheta {c_u}}}{\eta }} \right)}^i}} } \right)} \right]^{n + p}}dy.
\end{align}
\end{corollary}

Substituting $\varpi  = 0$ into \eqref{The asymptotic expression of the n-th user with ipSIC for CD-NOMA case1}, the asymptotic outage probability of the $n$-th user with pSIC for CD-NOMA is given by
\begin{align}\label{The asymptotic expression of the n-th user with pSIC for CD-NOMA case1}
P_{n,CD}^{pSIC,\infty } \approx \frac{{M!}}
{{\left( {M - n} \right)!n!}}{\left[ {\sum\limits_{u = 1}^U {\frac{{{b_u}}}
{{K!}}} {{\left( {\frac{{\beta {c_u}}}
{\eta }} \right)}^K}} \right]^n}.
\end{align}
\begin{remark}\label{remark2: the n-th user for CD-NOMA in case1}
Upon substituting \eqref{The asymptotic expression of the n-th user with ipSIC for CD-NOMA case1} and \eqref{The asymptotic expression of the n-th user with pSIC for CD-NOMA case1} into \eqref{The diversity order of COP}, the diversity orders of the $n$-th user with ipSIC/pSIC for CD-NOMA are zero and $nK$, respectively.
\end{remark}

\begin{corollary}\label{Corollary:the expression of asymptotic COP for the n-th user PD-NOMA case1}
For the special case with $K=1$, the asymptotic outage probability of the $n$-th user with ipSIC for PD-NOMA is given by
\begin{align}\label{The asymptotic expression of the n-th user with ipSIC for PD-NOMA case1}
 P_{n,PD}^{ipSIC,\infty } \approx &\frac{{{\phi _n}}}{{{\Omega _I}}}\sum\limits_{p = 0}^{M - n} {{
   {M - n}  \choose
   p
}\frac{{{{\left( { - 1} \right)}^p}}}{{n + p}}}  \nonumber \\
  &\times \int_0^\infty  {{e^{ - \frac{y}{{{\Omega _I}}}}}} {\left[ {\sum\limits_{u = 1}^U {{b_u}} \left( {1 - {e^{ - \frac{{y\vartheta {c_u}}}{\eta }}}} \right)} \right]^{n + p}}dy .
\end{align}
\end{corollary}
Substituting $\varpi  = 0$ into \eqref{The asymptotic expression of the n-th user with ipSIC for PD-NOMA case1}, the asymptotic outage probability of the $n$-th user with pSIC for PD-NOMA is given by
\begin{align}\label{The asymptotic expression of the n-th user with pSIC for PD-NOMA case1}
P_{n,PD}^{pSIC,\infty } \approx \frac{{M!}}
{{\left( {M - n} \right)!n!}}{\left[ {\sum\limits_{u = 0}^U {{b_u}} \left( {\frac{{\tau {c_u}}}
{\eta }} \right)} \right]^n}.
\end{align}
\begin{remark}\label{remark3: the n-th user for PD-NOMA in case1}
Upon substituting \eqref{The asymptotic expression of the n-th user with ipSIC for PD-NOMA case1} and \eqref{The asymptotic expression of the n-th user with pSIC for PD-NOMA case1} into \eqref{The diversity order of COP}, the diversity orders of the $n$-th user with ipSIC/pSIC for PD-NOMA are zero and $n$, respectively.
\end{remark}


\subsection{Throughput Analysis}\label{Throughput Analysis}
In this subsection, the system throughput of the unified NOMA framework is characterized in delay-limited transmission mode.
In this mode, the BS transmits information at a constant rate $R$, which is subject to the effect of outage probability. Hence the system throughput of CD/PD-NOMA with ipSIC/pSIC is given by
\begin{align}\label{Delay-limited Transmission for CD-NOMA}
R_\phi ^\psi  = \left( {1 - {P_{m,\phi }}} \right){R_n} + \left( {1 - P_{n,\phi }^\psi } \right){R_m},
\end{align}
where $\psi  \in \left( {ip{\rm{SIC}},pSIC} \right)$, $\phi  \in \left( {CD,PD} \right)$. ${P_{m,CD}}$ and ${P_{m,PD}}$ are given by
\eqref{CD-NOMA:the COP of far user} and \eqref{PD-NOMA:the COP of far user}, respectively. $P_{n,CD}^{ipSIC}$, $P_{n,CD}^{pSIC}$, $P_{n,PD}^{ipSIC}$ and $P_{n,PD}^{pSIC}$ are given by \eqref{CD-NOMA:the COP of near user for Case1 with ipSIC} and \eqref{CD-NOMA:the COP of near user for Case1 with pSIC}, \eqref{PD-NOMA :the COP of near user with ipSIC for Case1} and \eqref{PD-NOMA:the COP of near user with pSIC for Case1}, respectively.

\section{Numerical Results}\label{Numerical Results}
In this section, simulation results are presented to verify the analytical results derived in the above sections. In this unified framework considered, we assume the power allocation coefficients of a pair of users are $a_m=0.8$ and $a_n=0.2$, respectively.
The target rates are set to be $R_{{n}}=R_{{m}}=0.01 $ BPCU, where BPCU is short for bit per channel use. Setting the pathloss exponent  to be $\alpha=2$ and the system carrier frequency is equal to $1 $ GHz. The complexity-vs-accuracy tradeoff parameter is set to be $N=15$.
Without loss of generality, the OMA is selected to be a benchmark for comparison purposes.
Note that NOMA users with low target data rate can be applied to the IoT scenarios, which require low energy consumption, small packet service and so on.

\begin{figure}[t!]
    \begin{center}
        \includegraphics[width=3.2in,  height=2.4in]{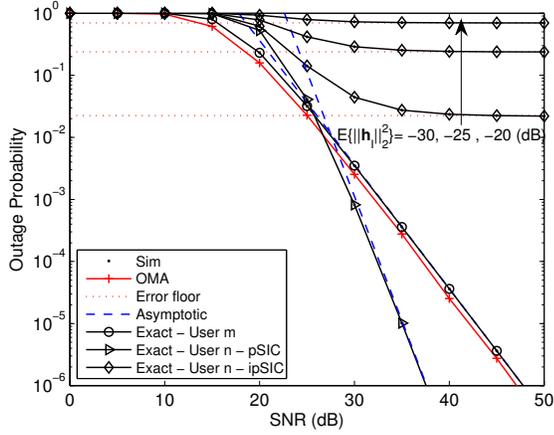}
        \caption{Outage probability versus the transmit SNR, with $K=2$, $M=3$, $n=2$, $m=1$, $R_D$ = 2 m, $R_n$ = $R_m$ = 0.01 BPCU.}
        \label{The_COP_pSIC_ipSIC_n_m_diff_LI}
    \end{center}
\end{figure}

Fig. \ref{The_COP_pSIC_ipSIC_n_m_diff_LI} plots the outage probability of a pair of NOMA users (the $m$-th and $n$-th user) versus the transmit SNR with pSIC/pSIC, where $K = 2$.
The exact analytical curves for the outage probability of the $m$-th user  is plotted according to \eqref{CD-NOMA:the COP of far user}. Furthermore, the exact analytical curves for the outage probability of the $n$-th user with ipSIC/pSIC are plotted based on \eqref{CD-NOMA:the COP of near user for Case1 with ipSIC} and \eqref{CD-NOMA:the COP of near user for Case1 with pSIC}, respectively.
Obviously, the exact outage probability curves match perfectly with the Monte Carlo simulations results. We observed that
the outage behavior of conventional OMA is inferior to the $n$-th user with pSIC and superior to the $m$-th user. This is due to the fact that NOMA is capable of providing better fairness since multiple users are served simultaneously, which is the same as the conclusions in \cite{Ding2014performance,Yuexinwei7812773}.

Additionally, as can be seen from Fig. \ref{The_COP_pSIC_ipSIC_n_m_diff_LI}, the dashed curves represent the asymptotic COP of the $m$-th user and $n$-th user with pSIC, which can be obtained by numerically evaluating \eqref{The expression of asymptotic COP for the m-th user with CD-NOMA} and \eqref{The asymptotic expression of the n-th user with pSIC for CD-NOMA case1}. One can observe that the asymptotic outage probabilities are approximated to the analytical results in the high SNR region.
The dotted curves represent the asymptotic outage probabilities of the $n$-th user with ipSIC, which are calculated from \eqref{The asymptotic expression of the n-th user with ipSIC for CD-NOMA case1}, respectively. It is shown that the outage performance of the $n$-th user converges to an error floor and obtain zero diversity order. Due to the influence of RI, the outage behavior of the $n$-th user with ipSIC is inferior to OMA. This is because that the RI signal from imperfect cancellation operation is the dominant impact factor. With the value of RI increasing from $-30$ dB to $-20$ dB, the outage behavior of the $n$-th user is becoming more worse and deteriorating. Hence the design of effective multiuser receiver algorithm is significant to improve the performance of NOMA networks in practical scenario.

\begin{figure}[t!]
    \begin{center}
        \includegraphics[width=3.2in,  height=2.4in]{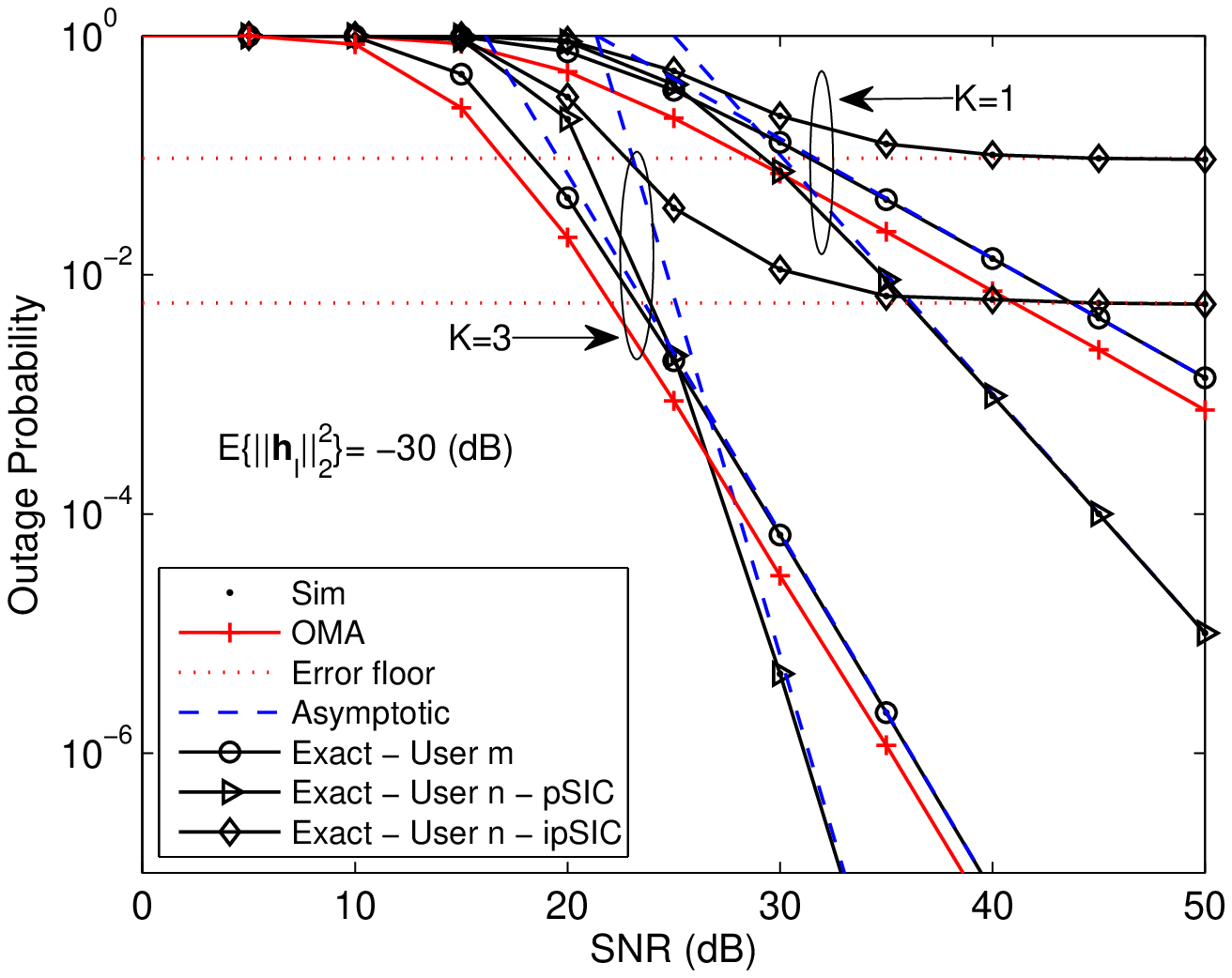}
        \caption{Outage probability versus the transmit SNR, with $M=3$, $n=2$, $m=1$, $R_n$ = $R_m$ = 0.01 BPCU.}
        \label{The_OP_m_n_pSIC_ipSIC_n_m_diff_K_add_OMA}
    \end{center}
\end{figure}

Fig. \ref{The_OP_m_n_pSIC_ipSIC_n_m_diff_K_add_OMA} plots the outage probability versus SNR with the different number of subcarriers (i.e., $K=3$ and $K=1$). For the special case with $K=1$, the unified framework of NOMA becomes PD-NOMA.
The exact outage probability curve of the $m$-th user for PD-NOMA is plotted according to \eqref{PD-NOMA:the COP of far user}. The exact outage probability curves of the $n$-th user with ipSIC/pSIC are given by Monte Carlo simulations and precisely match with the analytical expressions, which have been derived in \eqref{PD-NOMA :the COP of near user with ipSIC for Case1} and \eqref{PD-NOMA:the COP of near user with pSIC for Case1}, respectively. The asymptotic outage probabilities of this pair of users for PD-NOMA are also approximated with the analytical results in the high SNR region. We observe that CD-NOMA has a more steep slope and can provide better outage performance than PD-NOMA. This is due to the fact that CD-NOMA is capable of achieving the higher diversity orders.

\begin{figure}[t!]
    \begin{center}
        \includegraphics[width=3.2in,  height=2.4in]{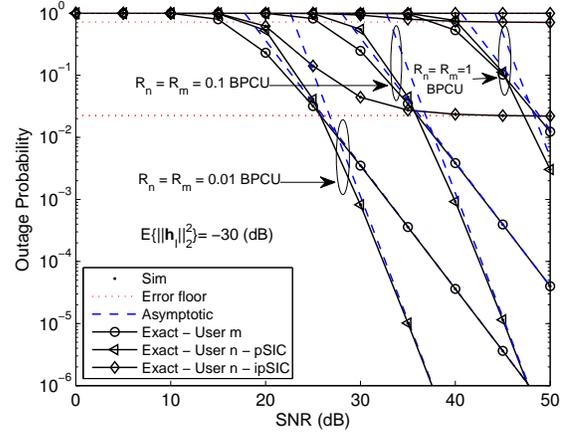}
      \caption{Outage probability versus the transmit SNR, with $M=3$, $n=2$, $m=1$, $K = 2$.}
        \label{The_OP_pSIC_ipSIC_n_m_diff_TR_new}
    \end{center}
\end{figure}

\begin{figure}[t!]
    \begin{center}
        \includegraphics[width=3.2in,  height=2.4in]{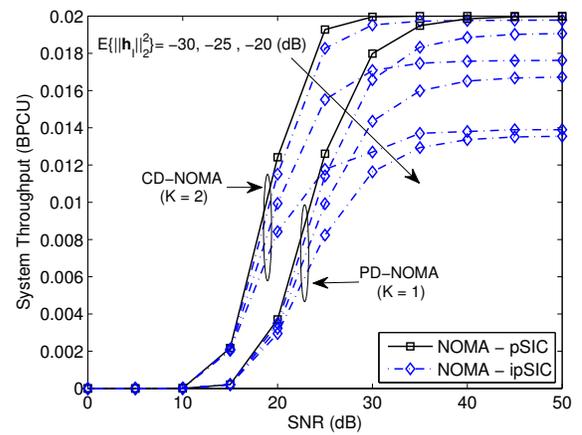}
        \caption{System throughput versus the transmit SNR, with $M=3$, $n=2$, $m=1$, $K = 2$.}
        \label{The_delay_limited_throughput_diff_LI}
    \end{center}
\end{figure}

Fig. \ref{The_OP_pSIC_ipSIC_n_m_diff_TR_new} plots the outage probabilities versus SNR for different user target rates. We observe that with increasing target rates, the lower outage probabilities are achieved. This is due to the fact that the achievable rates are directly related to the target SNRs. It is beneficial to detect the superposed signals for the selected user pairing with smaller target SNRs. It is worth pointing out that the impact of practical scenario parameter frequency dependent factor $\eta$ has been taken into account in the unified NOMA framework. Furthermore, the incorrect choice of $R_n$ and $R_m$ will lead to the improper outage behavior for the unified framework.

Fig. \ref{The_delay_limited_throughput_diff_LI} plots the system throughput versus SNR in the delay-limited transmission mode.
The solid black curves represent throughput of CD/PD-NOMA with ipSIC/pSIC, which can be obtained from \eqref{Delay-limited Transmission for CD-NOMA}. The dash-dotted blue curves represent throughput of CD-NOMA and PD-NOMA with ipSIC for the different values of RI.
We observe that CD-NOMA attains a higher throughput compared to PD-NOMA, since CD-NOMA has the smaller outage probabilities. This is due to that CD-NOMA is capable of attaining the larger diversity order than that of PD-NOMA.
Another observation is that increasing the values of RI from $-30$ dB to $-20$ dB will reduce the system throughput in high SNR region. This is because that CD/PD-NOMA converge to the error floors in the high SNR region.
\section{Conclusions}\label{Conclusion}
This paper has investigated the outage performance of a unified NOMA framework insightfully by invoking stochastic geometry.
The exact expressions for outage probability of a pair of users with ipSIC/pSIC have been derived.
It has been observed that the diversity orders of the $m$-th user for CD/PD-NOMA are $mK$ and $m$, respectively. However, due to the influence of RI, the diversity orders achieved by the $n$-th user with ipSIC are zeros for CD/PD-NOMA.
On the basis of analytical results, we observed that the outage behaviors of CD-NOMA is superior to that of PD-NOMA. Additionally,
the system throughput of CD/PD-NOMA with ipSIC/pSIC has been discussed in the delay-limited transmission mode.

\appendices
\section*{Appendix~A: Proof of Theorem \ref{Theorem:CD-NOMA:the COP of far user}} \label{Appendix:A}
\renewcommand{\theequation}{A.\arabic{equation}}
\setcounter{equation}{0}

The proof starts by assuming ${{{\bf{g}}_m}}$ and ${{{\bf{g}}_n}}$ have the same column weights for ${{\bf{G}}_{K \times M}}$.
That is to say that ${\left\| {diag\left( {{{\bf{h}}_m}} \right){{\bf{g}}_m}} \right\|_2^2}$ and ${\left\| {diag\left( {{{\bf{h}}_m}} \right){{\bf{g}}_n}} \right\|_2^2}$ follow the same distribution. Hence based on \eqref{SINR m}, the expression for outage probability of the $m$-th user is rewritten as
\begin{align}\label{The CDF of the m-th user}
{P_{m,CD}}   {\rm{ = {\text{Pr}}}}\left( {Z_m < \frac{{{{\varepsilon _m}}}}{{\rho \left( {{a_m} - \varepsilon _m {a_n}} \right)  }}}\buildrel \Delta \over = \tau \right),
\end{align}
where $Z_m = \left\| {diag\left( {{{\bf{h}}_m}} \right){{\bf{g}}_m}} \right\|_2^2 = \frac{\eta }{{1 + {d^\alpha }}}\sum\limits_{k = 1}^K {{{\left| {{g_{mk}}{{\tilde h}_{mk}}} \right|}^2}}$. It is observed that $Y = \sum\limits_{k = 1}^K {{{\left| {{g_{mk}}{{\tilde h}_{mk}}} \right|}^2}} $ is subject to a Gamma distribution with the parameters of $\left( {K,1} \right)$. The corresponding CDF of ${Y}$ is given by ${F_Y}\left( y \right) = 1 - {e^{ - y}}\sum\limits_{i = 0}^{K - 1} {\frac{{{y^i}}}{{i!}}}$.

In addition, on the basis of order statistics \cite{David2003Order}, the CDF of the sorted channel gains between the BS and users over $K$ subcarriers has a specific relationship with the unsorted channels, which can be expressed as follows:
\begin{align}\label{the CDF for Gamma distrbution}
{F_{Z_m}}\left( z \right) = {\phi _m}\sum\limits_{p = 0}^{M - m} {{
 M - m \choose
 p
 }\frac{{{{\left( { - 1} \right)}^p}}}{{m + p}}} {\left[ {{F_{\mathop Z\limits^ \sim  }}\left( z \right)} \right]^{m + p}},
\end{align}
where ${{F_{\mathop Z\limits^ \sim  }}\left( z \right)}$ denotes the CDF of unsorted channels for the $m$-th user.
Due to the assumption of homogeneous PPPs \cite{Yuanwei2017TWC} for randomly users and applying polar coordinate conversion, the CDF ${{F_{\mathop Z\limits^ \sim  }}\left( z \right)}$ is given by
\begin{align}\label{the CDF of Z for unsorted channels with stocastic}
{{F_{\mathop Z\limits^ \sim  }}\left( z \right)} = \frac{2}{{R_D^2}}\int_0^{{R_D}} {\left[ {1 - {e^{ - \frac{{z\left( {1 + {r^\alpha }} \right)}}{\eta }}}\sum\limits_{i = 0}^{K - 1} {\frac{1}{{i!}}{{\left( {\frac{{z\left( {1 + {r^\alpha }} \right)}}{\eta }} \right)}^i}} } \right]} rdr.
\end{align}
Obviously, it is difficult to obtain effective insights from the above integral. We employ the Gaussian-Chebyshev quadrature to provide an approximation of \eqref{the CDF of Z for unsorted channels with stocastic} and rewrite it as follows:
\begin{align}\label{the simplified CDF of Z for unsorted channels with stocastic}
{F_{\mathop Z\limits^ \sim  }}\left( z \right) \approx \sum\limits_{u = 1}^U {{b_u}} \left( {1 - {e^{ - \frac{{z {c_u}}}{\eta }}}\sum\limits_{i = 0}^{K - 1} {\frac{1}{{i!}}{{\left( {\frac{{z {c_u}}}{\eta }} \right)}^i}} } \right).
\end{align}
Substituting \eqref{the simplified CDF of Z for unsorted channels with stocastic} and \eqref{the CDF for Gamma distrbution} into \eqref{The CDF of the m-th user}, we can obtain \eqref{CD-NOMA:the COP of far user} and  complete the proof.

\appendices
\section*{Appendix~B: Proof of Theorem \ref{Theorem:CD-NOMA:the COP of near user for Case1 with ipSIC}} \label{Appendix:E}
\renewcommand{\theequation}{B.\arabic{equation}}
\setcounter{equation}{0}

Denote ${Z_n} = \left\| {diag\left( {{{\bf{h}}_n}} \right){{\bf{g}}_n}} \right\|_2^2 = \left\| {diag\left( {{{\bf{h}}_n}} \right){{\bf{g}}_m}} \right\|_2^2$ and ${Y_I} = \left\| {{{\bf{h}}_I}} \right\|_2^2$, respectively.
Substituting \eqref{The SINR expression at the n-th user to detect the m-th user} and \eqref{the SINR expression at n-th user to detect itself with SIC} into \eqref{CD-NOMA:the expression of COP for near user}, the COP of $P_{n,CD}^{ipSIC}$ can be expressed as
\begin{align}\label{The derived expression of the n-th user with ipSIC for the case1}
& P_{n,CD}^{ipSIC} = \underbrace {{\text{Pr}}\left( {\frac{{\rho {Z_n}{a_m}}}
{{\rho {Z_n}{a_n} + 1}} < {\varepsilon _m}} \right)}_{{J_1}} \hfill \nonumber \\
   & +  \underbrace {{\text{Pr}}\left( {\frac{{\rho {Z_n}{a_m}}}
{{\rho {Z_n}{a_n} + 1}} > {\varepsilon _m},\frac{{\rho {a_n}{Z_n}}}
{{\varpi \rho {Y_I} + 1}} < {\varepsilon _n}} \right)}_{{J_2}} \hfill ,
\end{align}
where ${J_1} = {F_{{Z_n}}}\left( \tau  \right)$, $\tau  = \frac{{{\varepsilon _m}}}{{\rho \left( {{a_m} - {\varepsilon _m}{a_n}} \right)}}$ with ${a_m} > {\varepsilon _m}{a_n}$ and $\varpi  = 1$.
Noting that ${Y_I}$ is also subjective to a Gamma distribution with the parameters of $\left( {K,{{\Omega _I}}} \right)$ and the corresponding PDF ${f_{{Y_I}}}$ is give by
\begin{align}\label{the PDF of interference h_I}
{f_{{Y_I}}}\left( y \right) = \frac{{{y^{K - 1}}{e^{ - \frac{y}{{{\Omega _I}}}}}}}{{\left( {K - 1} \right){\rm{!}} \Omega _I^K}}.
\end{align}

After some mathematical manipulations, ${J_2}$ is calculated as
\begin{align*}\label{The derived expression J_2 of the n-th user with ipSIC for the case1}
{J_2} = & {\text{Pr}}\left( {\tau  < {Z_n} < \vartheta {Y_I} + \beta } \right) \nonumber \\
\end{align*}
\begin{align}
    = & \underbrace {\int_0^\infty  {{f_{{Y_I}}}\left( y \right){F_{{Z_n}}}\left( {\vartheta y + \beta } \right)dy} }_{{J_3}} - {F_{{Z_n}}}\left( \tau  \right),
\end{align}
where $\beta {\text{ = }}\frac{{{\varepsilon _n}}}{{\rho {a_n}}}$ and $\vartheta {\text{ = }}\frac{{\varpi {\varepsilon _n}}}
{{{a_n}}}$.
Similar to the proving process of \eqref{The CDF of the m-th user}, based on \eqref{the PDF of interference h_I}, $J_3$ can be given by
\begin{align}\label{The derived expression J_3 of the n-th user with ipSIC for the case1}
&{J_3} \approx \frac{{{\phi _n}}}{{\left( {K - 1} \right)\Omega _I^K}}\sum\limits_{p = 0}^{M - n} {{
   {M - n}  \choose
   p
}\frac{{{{\left( { - 1} \right)}^p}}}{{n + p}}} \int_0^\infty  {{y^{K - 1}}{e^{ - \frac{y}{{{\Omega _I}}}}}}\nonumber  \\
&  \times {\left[ {\sum\limits_{u = 1}^U {{b_u}} \left( {1 - {e^{ - \frac{{{c_u}\left( {\vartheta y + \beta } \right)}}{\eta }}}\sum\limits_{i = 0}^{K - 1} {\frac{1}{{i!}}{{\left( {\frac{{\left( {\vartheta y + \beta } \right){c_u}}}{\eta }} \right)}^i}} } \right)} \right]^{n + p}}dy .
\end{align}
Substituting \eqref{The derived expression J_3 of the n-th user with ipSIC for the case1} and \eqref{The derived expression J_2 of the n-th user with ipSIC for the case1} into \eqref{The derived expression of the n-th user with ipSIC for the case1}, we can obtain \eqref{CD-NOMA:the COP of near user for Case1 with ipSIC} and complete the proof.

\bibliographystyle{IEEEtran}
\bibliography{mybib}

\end{document}